# RF Desense significance and its impact on the EVM at Signal Near the Noise Floor.

Rajesh Vedala, Harkirat Kaur, Palak Kapoor.
Microsoft, Surface RF DVT.

*Abstract*— Hardware impairments and system non-linearities impacting communication signal is one of key aspect for having harmonics and RF desense which overall causing the lower quality and integrity of the modulated signal, resulting in I/Q imbalance, further bit error and spectral efficiency degradation. This presentation outlines the RF Desense results, EVM Measurement and it impact at the almost noise floor with step size by 1 dB in QPSK at LTE Bands, Note for mmW 3GPP 38.521-2 clause 6.4.2.1 indicates single polarization.

## I. Background

EVM is one of the key metrics outlined to validate the TxRx characteristics of a communication device, the ability to receive and decode the signal at a moderate hardware imperfections and system's non-linearities (Electronic components tolerances-example LNA performance and output does not vary linearly in relation to the input of the system-example TxRx chain). It is a deviation measurement between the actual received signal vs ideal /reference symbol time (in-terms of both amplitude and phase error) of the given modulation method.

## II. Objective.

In the context of 3GPP and CTIA-PTCRB validation, the developed system—a compact, miniaturized integration of an antenna, RF cables, couplers, LNA, filters, HPA, and other components forming a complete RF chain— Operating in an electromagnetically (EM) radiated environment, while simultaneously facing thermal challenges from multiple subsystems—each running at its own clock frequency (conducted)—introduces a highly complex interference scenario. In addition to RF desense, these factors collectively result in a convoluted problem of problems, making system optimization a true challenge.

Such a combination of challenges can significantly impact receiver (Rx) performance, leading to potential bit errors, where a '0' may be misinterpreted as '1' or vice versa. Additionally, it can stress or overload symbol recovery mechanisms / autocorrelation in digital signal processing. This issue is particularly critical in high-order modulation schemes, such as 64-QAM in FR2 (3GPP spec single polarization) or in 256-QAM in sub6, where symbols are closely spaced, making it more susceptible to noise, interference, and system imperfections. These factors collectively degrade the overall communication device performance.

Besides improvements in HW, several digital filters helped overcome uncertainties such as fading, have improved SINR for a robust communication channel. Modulation and Coding and now with adaptive neural networks on sampling, leading to improved BLER, bit recovery.

Focus on this presentation:

1) RF Desense (LCD OFF vs ON) impact on LTE / Sub6 B20.
2) Example EVM Plot with calculation.
3) QPSK and QAM-64 performing at almost noise floor.

| LCD OFF Scenario | | | | LCD ON Scenario | | | | Desense Scenario (LCD ON = LCD ON - LCD OFF) | | | |
| --- | --- | --- | --- | --- | --- | --- | --- | --- | --- | --- | --- |
| Band | Freq (MHz) | Rx0 (dBm) | Rx1 (dBm) | Band | Freq (MHz) | Rx0 (dBm) | Rx1 (dBm) | Band | Freq (MHz) | Rx0 (dBm) | Rx1 (dBm) |
| LTE_B20 | 810.2 | -97.35 | -99.41 | LTE_B20 | 810.2 | -94.06 | -94.22 | LTE_B20 | 810.2 | 3.29 | 5.19 |
| LTE_B20 | 810.5 | -97.34 | -99.24 | LTE_B20 | 810.5 | -93.84 | -94.15 | LTE_B20 | 810.5 | 3.5 | 5.09 |
| LTE_B20 | 810.8 | -97.33 | -99.21 | LTE_B20 | 810.8 | -94.05 | -93.95 | LTE_B20 | 810.8 | 3.28 | 5.26 |
| LTE_B20 | 811.1 | -97.27 | -99.15 | LTE_B20 | 811.1 | -93.76 | -94.02 | LTE_B20 | 811.1 | 3.51 | 5.13 |
| LTE_B20 | 811.4 | -97.3 | -99.19 | LTE_B20 | 811.4 | -93.92 | -93.92 | LTE_B20 | 811.4 | 3.38 | 5.27 |
| LTE_B20 | 811.7 | -97.41 | -99.27 | LTE_B20 | 811.7 | -93.85 | -93.97 | LTE_B20 | 811.7 | 3.56 | 5.3 |
| LTE_B20 | 812 | -97.53 | -99.34 | LTE_B20 | 812 | -93.91 | -93.83 | LTE_B20 | 812 | 3.62 | 5.51 |
| LTE_B20 | 812.3 | -97.55 | -99.4 | LTE_B20 | 812.3 | -94.15 | -93.94 | LTE_B20 | 812.3 | 3.4 | 5.46 |
| LTE_B20 | 812.6 | -97.58 | -99.3 | LTE_B20 | 812.6 | -94.25 | -93.67 | LTE_B20 | 812.6 | 3.33 | 5.63 |

Table1.0

### 1) RF Dense (LCD ON) impact on LTE / Sub6 signal

Measuring the RF sentivity across the operating frequencies for multiple Rx Antennas and outlining its values as above Table 1.0, further applying the desense scenario such as LCD ON in this case, refer above Table 1.0. i.e., measure Rx sensitivity power level for Rx Primary and Rx Diversity Antenna chains during LCD OFF, repeat the same process with LCD ON and subtract LCD ON – LCD OFF = Desense value (Aggressor vs Victim scenario), example when 2dB is increased during LCD-ON that means the LCD-ON is causing an impact by 2dB to the Rx chains.

For instance, if the measured sensitivity degrades by 2 dB when the LCD is turned ON, it is denoting the LCD display (LCD ON) acts as an aggressor, causing a 2 dB reduction (example at 800MHz it is approx 2.4dB for Rx Div antenna) in the Rx chain performance, with the antennas functioning as the victim in this interference scenario, refer Figure - (A).

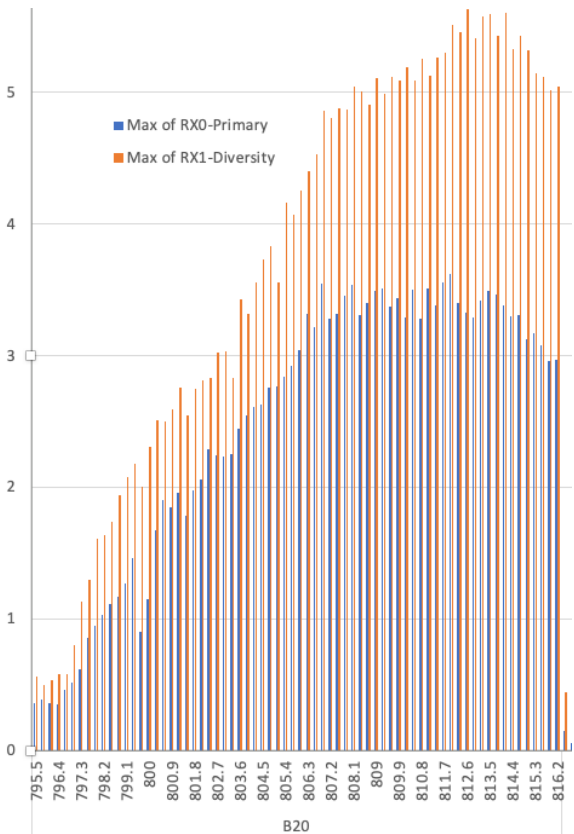

Figure – (A).

### 2) Example EVM calculation using QPSK

Measures the distortion and quality of the received signal, the calculation as below with ideal constellation points represented by in-phase (I) and quadrature (Q) components as in sample code – E and same measured constellation points as in sample code – E, refer Figure – (B) for matlab plot of ideal symbols and the sample code from matlab as Function 1. The calculation as

$$EVM_{values} = (I_{meas} - I_{ref})^2 + (Q_{meas} - Q_{ref})^2$$

$$EVM_{percentage} = 2 EVM_{rms} \times 100$$

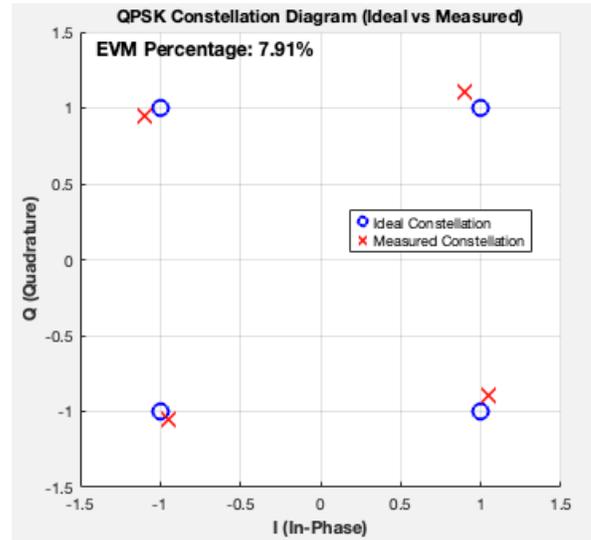

Figure – (B).

Function 1 EVM calculation

```
% Ideal bits (reference constellation points)
Iref = [1 -1 -1 1];
Qref = [1 1 -1 -1];

% Measured bits (measured constellation points)
Imeas = [0.9 -1.1 -0.95 1.05];
Qmeas = [1.1 0.95 -1.05 -0.9];

% Calculate EVM
evm_values = sqrt((Imeas - Iref).^2 + (Qmeas - Qref).^2);
evm_rms = sqrt(mean(evm_values.^2));

% EVM in percentage
evm_percentage = evm_rms / sqrt(2) * 100;
```

### 3) Impact on EVM for every 1dB step at Rx Senstivity.

The impact on Error Vector Magnitude (EVM) is analyzed for each 1 dB step change in Rx sensitivity. By applying the previously obtained Rx sensitivity measurements, the results are correlated with the theoretical receiver degradation occurs when the low-noise amplifier (LNA) enters saturation, causing nonlinear signal amplification. This leads to increased noise, reduced symbol correlation, and higher Error Vector Magnitude (EVM) due to HW impairments as a cross-check. This relation can be observed in the below example for the captured EVM plots at -80dBm for 10MHz Bandwidth at LTE B20 (approx. 820 MHz) with 1dB increment step size (i.e., -80 to -84dBm) 4 snapshots. The symbols are spread-out as the signal level reaches almost noise floor thus recovery of symbols gets difficult stressing bit recovery mechanism. refer Figure - (C).

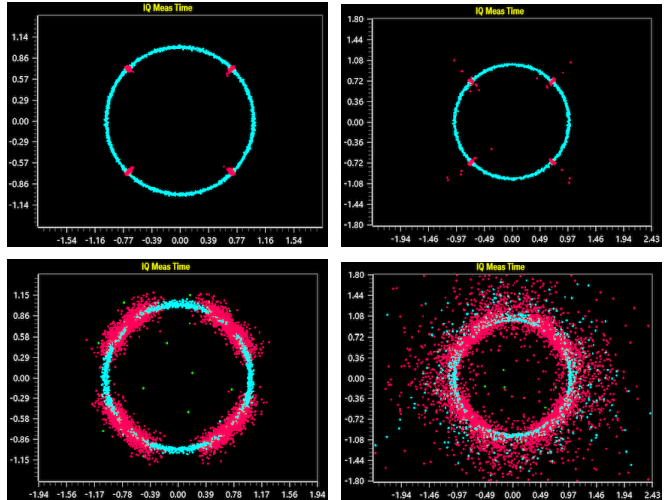

Figure – (C).

Similarly, for FR2 mmW n260 Hi Ch, 100MHz BW, 120KHz SCS, Rx power at approx. -70dBm, 16-QAM below are the EVM plots, where the EVM error rate % has increased more rapidly along with beam switching, which was instantaneous due to migrating to improved Beam ID., refer Figure (D).

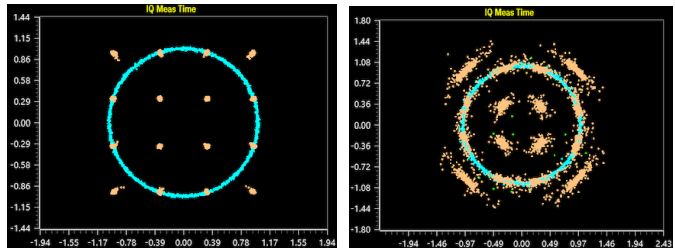

Figure – (D).

## III. FINDINGS.

This presentation provides an insight on step-by-step guidance in understanding the importance and its impact on EVM performance which is dependent on RF Desense on how clean the HW is free from noise and harmonics.

## ACKNOWLEDGMENT

Sincere thanks to my boss, my colleagues at Surface Engineering and continuous support, technical guidance, along with help from Infrastructure teams (Keysight, Anritsu) conducting multiple experimental tests and validation of obtained results in collaboration with product development, validation, test methodologies.